\documentclass[twocolumn,showpacs,preprintnumbers,amsmath,amssymb,prd,aps]{revtex4-1}

\usepackage{graphicx}
\usepackage{dcolumn}
\usepackage{bm}
\usepackage{amsmath}
\usepackage{relsize}

\begin{document}

\title{Constraints on the origin of the ultra-high energy cosmic-rays
   using cosmic diffuse neutrino flux limits: An analytical approach}

\author{Shigeru Yoshida}
\thanks{syoshida@hepburn.s.chiba-u.ac.jp (S.~Yoshida)}
\author{Aya Ishihara}
\thanks{aya@hepburn.s.chiba-u.ac.jp (A.~Ishihara)}

\affiliation{%
Department of Physics, Graduate School of Science, Chiba University, Chiba 263-8522, Japan
}%

\date{\today}

\begin{abstract}
Astrophysical neutrinos are expected to be produced in the interactions
of ultra-high energy cosmic-rays with surrounding photons. 
The fluxes of the astrophysical neutrinos are highly dependent on
the characteristics of the cosmic-ray sources, such as their cosmological
distributions. We study possible constraints on the properties of cosmic-ray sources in a
model-independent way using experimentally obtained diffuse neutrino flux above 100~PeV.
The semi-analytic formula is derived to estimate the cosmogenic neutrino fluxes as functions of source
evolution parameter and source extension in redshift.
The obtained formula converts the
upper-limits on the neutrino fluxes into the constraints on the
cosmic-ray sources. It is found that the recently obtained upper-limit
on the cosmogenic neutrinos by IceCube constrains the scenarios with strongly evolving 
ultra-high energy cosmic-ray sources, and the future limits from an 1~km$^3$ scale detector
are able to further constrain the ultra-high energy cosmic-rays
sources with evolutions comparable to the cosmic star formation rate.
\end{abstract}

\maketitle

\section{\label{sec:intro} Introduction}

The origin of the ultra-high energy cosmic-rays (UHECRs) has been a long-standing
important question in astrophysics. While the observations by Auger~\cite{auger07,auger08} and HiRes~\cite{hires04}
indicate that cosmic-rays with energies above $\sim$$10^{18.5}$eV are of extra-galactic origin, 
identification of astronomical objects responsible for the UHECR emission has not been achieved.
Neutrinos, secondary produced by UHECR nucleons, 
are expected to provide the direct information on the UHECR origin, since a neutrino 
penetrates over cosmological distance without being deflected
by cosmic magnetic field nor absorbed by the photon field.
In particular, the intensity of the 
``cosmogenic'' neutrinos~\cite{berezinsky69} 
produced by the collisions of UHECR nucleons with the cosmic microwave 
background (CMB) photon via photo-produced $\pi$ meson decay as
$\pi^{\pm}\to\mu^{\pm}\nu_\mu\to e^{\pm}\nu_e\nu_\mu$, 
known as the Greisen-Zatsepin-Kuzmin (GZK) mechanism~\cite{GZK},
indicates redshift distributions of the parent UHECR sources~\cite{yoshida93,yoshida97} in the Universe.
The source distributions derived from the cosmogenic neutrino 
intensities can then be compared with distributions of known classes of 
the astronomical objects possibly responsible for the UHECR emissions.
Therefore reliable extractions of the UHECR source distribution function (s.d.f.) 
is one of key issues in cosmogenic diffusive neutrino searches.
Constraints on the sources of UHECRs derived from the measurements or upper-limits of the 
ultra-high energy neutrino flux are highly complimentary to the constraints from the 
diffuse photon flux~\cite{fermilimit,ahlers2010}, because the former does not rely on uncertain estimation 
of extra-galactic background light (EBL).

In this work, we develop a method to bound the UHECR source evolution and its redshift dependence
in a comprehensive way without introducing specific astronomical models.
We derive an analytical formula to calculate intensities of the neutrinos produced by the GZK mechanism 
in the range between 100~PeV and 10~EeV.
%
Using the formula, we extract the relation among the neutrino intensity and the UHECR s.d.f.\ 
parameters. The use of the analytical formula allows 
us to calculate neutrino intensities in the full phase space 
of the source evolution parameters without an intensive computational task.
The analytical formula can also be used as a practical tool to approximately calculate
cosmogenic GZK neutrino intensity with given UHECR s.d.f., for example, for the performance 
studies of the future detectors such as KM3NET~\cite{km3net}.
Finally we present model independent constraints on the UHECR sources using the obtained formula 
with the upper-limit~\cite{icecubeEHE2011} and the future sensitivity~\cite{aya2011} 
on the cosmogenic neutrino detection by the IceCube neutrino observatory~\cite{icecube}. 

The standard cosmology with $H_0 \simeq 73.5$ km $\sec^{-1}$ Mpc$^{-1}$,
$\Omega_M = 0.3$, and $\Omega_{\Lambda}=0.7$~\cite{PD2010} is assumed throughout the paper.

\section{\label{sec:function} Analytical formula for estimating cosmogenic $\nu$ intensity}

The neutrino flux per unit energy, $dJ_{\nu}/dE_{\nu}$, is generally written as,
\begin{eqnarray}
{dJ_{\nu} \over dE_{\nu}}&& = n_0 c \int\limits^{z_{max}}_0 
\psi(z_{s}) \left|{dt \over dz} (z_s)\right| dz_{s} \label{eq:general_t}\\
 \int\limits^{z_{s}}_0&& \left|{dt \over dz} (z_\nu)\right| dz_\nu
\int\limits^\infty_{E_\nu}{dN_{p\to\nu}\over dE^{g}_{\nu}dt^g}(z_{\nu},z_{s})\delta(E_{\nu}^{g}\!-\!(1+z_\nu)E_{\nu}) dE_\nu^{g}. \nonumber\\
\nonumber
\end{eqnarray}
The first integral represents the total contribution of UHECR sources in the redshift up to $z_{max}$, 
where $z_{max}$ is the maximum redshift of the UHECR source distribution or, in other words, 
the time of the first UHECR emission in the Universe.
$\psi(z_{s})$ represents the cosmic evolution of the spectral emission rate per co-moving volume
and $n_0$ is the number density of UHECR sources at the present universe.
The relation between time and redshift is given as 
$|\frac{dt}{dz}(z_s)| \equiv |\frac{dt_s}{dz_s}| = [H_0(1+z_s)\sqrt{\Omega_M(1+z_s)^3+\Omega_\Lambda}]^{-1}$. 
The second integral calculates the total neutrino flux expected from 
a single UHECR source at redshift $z_{s}$ generated via UHECR interactions at redshift $z_{\nu} (\leq z_{s})$. 
$dN_{p\to\nu}/dE^{g}_{\nu}dt^g$ is the yield of generated neutrinos with
energy of $E^{g}_{\nu}$ per unit time in the UHECR laboratory frame (the CMB rest frame).
Suffixes $s$ and $g$ represent the quantities at the positions of UHECR sources 
and neutrino generation, respectively.
The delta function indicates the neutrino energy loss due to the expansion of the universe.

The neutrino yield, $dN_{p\to\nu}/dE^{g}_{\nu}dt^g$, at redshift $z_\nu$ by the GZK mechanism 
is expressed by a convolution of the UHECR intensity
from a source at $z_{s}$, the CMB photon density, and the photo-pion interaction kinematics as,
\begin{eqnarray}
{dN_{p\to\nu}\over dE^{g}_\nu dt^g}&=& \int dE_{\rm CR} {dN_{\rm CR}\over dE_{\rm CR}}(z_{s}, z_\nu) \nonumber\\
&& c \int ds \int dE_{\pi}{d\sigma_{\gamma p} \over dE_\pi}
{d\rho_{\pi\to\nu}\over dE_\nu^{g}}{dn_\gamma\over ds},
\label{eq:neut_yield}
\end{eqnarray}
where $dN_{\rm CR}/dE_{\rm CR}$ is number of the UHECRs per unit time and energy
at the redshift $z_\nu$ originating from a source at $z_s$,
and $s$ is the Lorentz-invariant Mandelstam variable, the square of invariant mass 
of the cosmic-ray nucleon and the target CMB photon.
$\sigma_{\gamma p}$ is the photo-pion production cross section,
$d\rho_{\pi\to\nu}/dE_\nu^{g}$ is the energy distribution of neutrinos from the photo-produced
pion, and $dn_\gamma/ds$ is the CMB photon number density in the UHECR frame
per unit $s$.

We introduce following approximations to simplify the calculations: 
1) the contribution of UHECR colliding with IR/O background is negligible
and only the contribution of photo-pion production cross-section from $\Delta$-resonance is considered
in collisions of UHECRs and CMB photons, and 
2) the kinematics of the photo-pion production is represented by a single pion production.
The first approximation allows the photon number density $dn_\gamma/ds$ 
to be analytically obtained with the modification to the black-body distribution~\cite{yoshida93}.
The contribution of neutrinos induced by UHECR interactions with IR/O 
becomes sizable only in energy region below 100~PeV~\cite{kotera2010} while
the effect is small in the higher energy region.
Similarly the neutrinos from photo-produced pions outside
the $\Delta$-resonance are mostly visible only in the lower energy range
below 100~PeV~\cite{ESS}, and the single pion production is the most dominant channel in 
the $\Delta$-resonance. 
The $\Delta$-resonance approximation simplifies the integral on $s$ in Eq.~\ref{eq:neut_yield} 
to a multiplication of the integrand at $s=s_R(\simeq1.5\ {\rm GeV}^2)$, 
where $s_R$ is the Lorentz-invariant Mandelstam variable at the $\Delta$-resonance, 
and  $\Delta s_R (\simeq 0.6\ {\rm GeV}^2)$, the width of the $\Delta$-resonance.
%
The 2nd approximation then gives~\cite{gaisser90},
\begin{equation}
{d\rho_{\pi\to\nu}\over dE_\nu^{g}} \simeq {1\over E_\pi}{3\over 1-r_\pi},
\label{eq:pi_decay}
\end{equation}
where $r_\pi=m_\mu^2/m_\pi^2\simeq 0.57$ is the muon-to-pion mass squared ratio.
The factor three arises from the fact that three neutrinos are produced 
from the $\pi$ meson and $\mu$ lepton decay chain.
The allowed range of $E_\nu^{g}$ due to the kinematics is given by,
\begin{equation}
0 \leq {E_\nu^{g}\over E_\pi} \leq 1-r_\pi,
\label{eq:neut_energy_range}
\end{equation}
where neutrino mass is neglected. With a good approximation
that a single pion is isotropically emitted in the center-of-momentum frame, one obtains
\begin{eqnarray}
{d\sigma_{\gamma p}\over dE_\pi}&=& {1\over E_{\rm CR}}{d\sigma_{\gamma p}\over dx_\pi}\\\nonumber
 &\simeq& {1\over E_{\rm CR}}{\sigma_{\gamma p}\over x^+-x^-},
\end{eqnarray}
where $x_\pi\equiv E_\pi/E_{\rm CR}$ is the relative energy of emitted pion normalized
by the parent proton energy $E_{\rm CR}$.  $x^{\pm}$ are the maximal and minimal bound
of $x_\pi$ due to the kinematics and given by
\begin{equation}
x^\pm = {s+m_\pi^2-m_p^2\over 2s}\pm {\sqrt{(s+m_{\pi}^2-m_p^2)^2-4sm_\pi^2}\over 2s},
\label{eq:eta_bound}
\end{equation}
where $m_p$ is the proton mass. 

Then we obtain the neutrino yield, Eq.~\ref{eq:neut_yield}, expressed
as an analytical function with only a single energy integral;
{\small
\begin{eqnarray}
{dN_{p\to\nu}\over dE^{g}_\nu dt^d}&\simeq& 
{k_{\rm B}T(1+z_\nu)\over 8\pi^2\hbar^3c^2}(s_R-m_p^2)\sigma_{\gamma p}^{R}\nonumber\\
&& {s_R\Delta s_R\over \sqrt{(s_R+m_{\pi}^2-m_p^2)^2-4s_Rm_\pi^2}} {3\over 1-r_\pi}\nonumber\\
&&\int dE_{\rm CR} {1\over E_{\rm CR}^3}{dN_{\rm CR}\over dE_{\rm CR}}\ln\left({x^+_R\over\xi_R}\right)\nonumber\\
&& \left\{-\ln\left(1-e^{-{E_{\Delta}\over (1+z_\nu)E_{\rm CR}}}\right)\right\}.
\label{eq:neut_yield_cmb}
\end{eqnarray}
}
Here $k_{\rm B}$ is the Boltzmann constant, $T$ is present temperature of the CMB. 
$E_{\Delta}\equiv(s_R-m_p^2)/4k_BT$  corresponds to the energy of UHECR protons
colliding via $\Delta$-resonance at the present universe. 
Suffix $R$ denotes the values at the $\Delta$-resonance in the photo-pion reaction.
For example, $\sigma_{\gamma p}^{R}=2.1\times 10^{-28} {\rm cm}^2$ represents the photo-pion production
cross-section of channel $\gamma p \to n\pi^+$ at the $\Delta$ resonance 
and $x^\pm_R$ is given by Eq.~\ref{eq:eta_bound} with $s=s_R$.

The parameter $\xi_R$ reflects the kinematics bounds, 
Eqs.~\ref{eq:neut_energy_range} and \ref{eq:eta_bound}, 
and is defined by
\begin{equation}
\xi_R = \begin{cases}
x^-_R                         & E_\nu^g \leq (1-r_\pi)x_R^{-} E_{\rm CR},\\
{E_\nu^g\over (1-r_\pi)E_{\rm CR}} & \text{otherwise}.
\end{cases}
\label{eq:xi_range}
\end{equation}

$dN_{\rm CR}/dE_{\rm CR}$ is 
calculated by the energy loss formula with the Continuous Energy Loss (CEL) approximation~\cite{CEL}
represented by
\begin{equation}
-{dE_{\rm CR} \over cdt} = {(1+z)^3\over\lambda_{\rm GZK}\left(E_{\rm CR}(1+z)\right)}E_{\rm CR},
\label{eq:gzk_CEL}
\end{equation}
where $\lambda_{\rm GZK}$ is the energy attenuation length governed by the GZK mechanism, 
mainly due to the photo-pion production of UHECRs and the CMB. The factor $(1+z)^3$ accounts for the increase
of CMB photon number density with redshift $z$. 

Here we introduce the final approximation that the energy attenuation length 
of UHECR by the GZK mechanism, $\lambda_{\rm GZK}$, is constant with energies above $E_{\rm GZK}\equiv 10^{20}$~eV. 
While $\lambda_{\rm GZK}$ rapidly decreases with cosmic-ray energy increase, it turns to be a slight increase or constant above $\sim3\times 10^{20}$~eV for $z_{\nu} \sim 0$.
Neutrinos from $z_{\nu}\gtrsim 1$ are dominant contribution to the cosmogenic neutrino intensity at earth and
the turnover energy are shifted to lower energies $\lesssim E_{\rm GZK}$ due to the redshift effects for the universe of $z_{\nu}\gtrsim 1$~\cite{takami09}.
Therefore this approximation reasonably describes the UHECR energy loss profile to calculate the neutrino yield.
Assuming the primary UHECR spectrum from a source at $z_s$ follows the power law described by
$dN_{\rm CR} (z_s, z_s)/dE_{\rm CR} = \kappa_{\rm CR}(E_{\rm CR}/E_{\rm GZK})^{-\alpha}$ up to $E_{\rm max}$, the maximal injected energy from a source,
then the $dN_{\rm CR} (z_s, z_\nu)/dE_{\rm CR}$ is analytically obtained by,
{\small
\begin{eqnarray}
{dN_{\rm CR}\over dE_{\rm CR}}(z_s,z_{\nu})&=&\kappa_{\rm CR}({E_{\rm CR}\over E_{\rm GZK}})^{-\alpha}\nonumber\\
 && e^{-(\alpha-1){c\over\lambda_{\rm GZK}H_0}{2\over 3\Omega_M}\left\{f(z_{s})-f(z_\nu)\right\}},
\label{eq:UHECR_spectrum}
\end{eqnarray}
}
where $f(z) \equiv \sqrt{\Omega_M(1+z)^3+\Omega_{\Lambda}}$ and $\kappa_{\rm CR}$ is a normalization constant.
With Eq.~\ref{eq:UHECR_spectrum}, the energy integral on $E_{\rm CR}$ in Eq.~\ref{eq:neut_yield_cmb} 
becomes an addition of integrals in the forms 
of $\int dy y^{-(\alpha+3)}\ln(1-e^{-1/y})$ and $\int dy y^{-(\alpha+3)}\ln(1-e^{-1/y})\ln{y}$.
An asymptotic approximation with numerical constants is found to provide
approximate solutions of these integrals.
The final formula of the neutrino yield is then obtained as,
{\small
\begin{eqnarray}
{dN_{p\to\nu}\over dE^{g}_\nu dt^g}&=&\kappa_{\rm CR}
{k_{\rm B}T\over 8\pi^2\hbar^3c^2}{(s_R-m_p^2)\over E_{\rm GZK}^2}\left({E_{\Delta}\over E_{\rm GZK}}\right)^{-(\alpha+2)}\nonumber\\
&& \sigma_{\gamma p}^{R}{s_R\Delta s_R\over \sqrt{(s_R+m_{\pi}^2-m_p^2)^2-4s_Rm_\pi^2}}{3\over 1-r_\pi}\nonumber\\
&& (1+z_\nu)^{\alpha+3}e^{-(\alpha-1){c\over\lambda_{\rm GZK}H_0}{2\over 3\Omega_M}\left\{f(z_{s})-f(z_\nu)\right\}}\nonumber\\
&& \left\{\epsilon_0x_0^{-(\alpha+1)}e^{-2}\ln\left({x_R^+\over x_R^-}\right)+\epsilon_1x_1^{-(\alpha+3)}e^{-1/x_1}e^{-2}\right. \nonumber\\
&& \left.\ln\left(x_1{E_{\Delta}\over E_\nu^{g}(1+z_\nu)}x_R^-(1-r_\pi)\right)\right\},\label{eq:neut_yield_final} \\
\nonumber
\end{eqnarray}
}
where $x_0=0.275$, and $x_1=0.16$ are the empirically determined numerical constants.
$\epsilon_0$ and $\epsilon_1$ are either unity or null, depending on neutrino energy.
These are consequences of the kinematics bound for pions and neutrinos (Eq.~\ref{eq:xi_range}), and given by
{\small
\begin{equation}
\epsilon_0=\begin{cases}
1& E_\nu^g(1+z_\nu)\leq x_1E_{\Delta}x_R^+(1-r_\pi), \\
0& \text{otherwise}.
\end{cases}
\label{eq:epsilon_0}
\end{equation}
}
and
{\small
\begin{equation}
%
\epsilon_1\! =\! \begin{cases}
0 & E_\nu^g(1+z_\nu)\leq x_1E_{\Delta}x_R^-(1-r_\pi),\\
1 & x_1E_{\Delta}x_R^-(1\!-\!r_\pi){\leq E_\nu^g(1\!+\!z_\nu) \leq} x_1E_{\Delta}x_R^+(1\!-\!r_\pi),\\
0 & \text{otherwise}.
\end{cases}
\label{eq:epsilon_1}
\end{equation}}
One can also find that $x_1E_{\Delta}x_R^\pm(1-r_\pi)$
in these equations represents the effective energy of neutrinos from decay of the pions 
of which energies are kinematically allowed energy maximum ($x_R^+$) 
and minimum ($x_R^-$) from $\Delta$-resonance in the $\gamma p$ collision. 
$E_\nu^{g}(1+z_\nu)=E_\nu(1+z_\nu)^2$ factor
reflects the redshift dependence of the CMB temperature and
the redshift energy loss of neutrinos at $z_\nu$.

Eq.~\ref{eq:general_t} with the formula Eq.~\ref{eq:neut_yield_final} 
finally gives the cosmogenic neutrino flux with double integrals of redshift $z_{s}$ and $z_{\nu}$.
The $z_\nu$ integral is analytically solvable neglecting $O((\lambda_{\rm GZK}H_0/c)^2)$ or higher order terms as the energy attenuation length
is much shorter than the cosmological time dimension. Then the final form
of the cosmogenic neutrino intensity is obtained as,
{\small
\begin{eqnarray}
{dJ_{\nu}\over dE_{\nu}}&=&(\alpha-1)F_{\rm CR}{c\over H_0}
{k_{\rm B}T\over 8\pi^2\hbar^3c^3}\nonumber\\
&& {(s_R-m_p^2)\over E_{\rm GZK}^3}\left({E_{\Delta}\over E_{\rm GZK}}\right)^{-(\alpha+2)}\nonumber\\
&& \sigma_{\gamma p}^{R}{s_R\Delta s_R\over \sqrt{(s_R+m_{\pi}^2-m_p^2)^2-4s_Rm_\pi^2}}
{3\over 1-r_\pi}\zeta.
\label{eq:gzk_final}
\end{eqnarray}
}
Here $F_{\rm CR}$ represents the UHECR intensity above $E_{\rm GZK}$ and
described by
\begin{eqnarray}
F_{\rm CR}& = & \int\limits^{E_{\rm max}}_{E_{\rm GZK}}dE_{\rm CR}
n_0 c \int\limits^{z_{max}}_0 \psi(z_{s}) \left|{dt \over dz} (z_s)\right| dz_{s}{dN_{\rm CR}\over dE_{\rm CR}}(z_s,0) \nonumber\\
 &\simeq& n_0\kappa_{\rm CR}E_{\rm GZK}\lambda_{\rm GZK}/(\alpha-1)^2
\label{eq:F_CR}
\end{eqnarray}
assuming that $E_{\rm max}\gg E_{\rm GZK}$. $F_{\rm CR}$ gives the normalization
of the neutrino flux in the present formulation in Eq.~\ref{eq:gzk_final}.
It can be estimated by the observational data 
for actual calculation.

$\zeta$ in Eq.~\ref{eq:gzk_final} is the term which account to the redshift dependence and given by,
{\small
\begin{eqnarray}
\zeta &=& \int\limits^{z_{max}}_0 dz_{s}
{(1+z_{s})^{(m+\alpha-1)}\over\sqrt{\Omega_M(1+z_{s})^3+\Omega_{\Lambda}}}
 \left\{\epsilon_0x_0^{-(\alpha+1)}e^{-2}\ln\left({x_R^+\over x_R^-}\right)+\right. \nonumber\\
 && \left.\epsilon_1x_1^{-(\alpha+3)}e^{-1/x_1}e^{-2}\right. \nonumber\\
 && \left.\ln\left(x_1{E_{\Delta}\over E_\nu(1+z_{s})^2}x_R^-(1-r_\pi)\right)\right\},
\label{eq:gzk_redshift}
\end{eqnarray}
}
where $\epsilon_0$ and $\epsilon_1$ are obtained by Eq.~\ref{eq:epsilon_0} and \ref{eq:epsilon_1} respectively, 
replacing $z_\nu$ by $z_{s}$. The cosmic evolution function
$\psi(z_{s})$ is now parametrized as $(1+z_{s})^m$ such that the parameter $m$ represents the ``scale''
of the cosmic evolution often used in the literature~\cite{yoshida03}.
The integral on $z_s$ in Eq.~\ref{eq:gzk_redshift} is analytically solvable 
when we use the fact that $\sqrt{\Omega_M(1+z_{s})^3+\Omega_\Lambda}\gg \Omega_\Lambda$
in most of the integral range. 
Finally we obtain the final form of the red shift dependent part of the analytical formula $\zeta$ as,
{\small
\begin{eqnarray}
\zeta &=& e^{-2}{1\over \gamma_m}\Omega_M^{-{\alpha+m\over 3}} \nonumber\\
&&\left\{(\Omega_M(1+z_{up})^3+\Omega_{\Lambda})^{\gamma_m\over 3}
\left(x_0^{-(\alpha+1)}\ln{\left({x^{+}_R\over x^{-}_R}\right)}+ \right.\right. \nonumber\\
&& \left.\left. x_1^{-(\alpha+3)}e^{-{1\over x_1}}
\left[\ln\left({x_1{E_{\Delta}\over E_\nu(1+z_{up})^2}x^{-}_R(1-r_\pi)}\right)\right.\right.\right. \nonumber\\
&& \left.\left.\left. +{2\over \gamma_m}\right]\right) -(\Omega_M+\Omega_{\Lambda})^{\gamma_m\over 3}x_0^{-(\alpha+1)}
\ln{\left({x^{+}_R\over x^{-}_R}\right)}\right.\nonumber\\
&& \left. -(\Omega_M(1+z_{down})^3+\Omega_{\Lambda})^{\gamma_m\over 3}
x_1^{-(\alpha+3)}e^{-{1\over x_1}}\right.\nonumber\\
&& \left. \left[\ln\left({x_1{E_{\Delta}\over E_\nu(1+z_{down})^2}x^{-}_R(1-r_\pi)}\right) +
{2\over \gamma_m}\right]\right\}\, \label{eq:gzk_redshift_final}\\
\nonumber
\end{eqnarray}
}
where $\gamma_m \equiv (\alpha+m)-3/2$. 
$z_{up}$ and $z_{down}$ are the maximum and minimum bounds of the redshifts, respectively.
These redshift bounds are associated with $z_{max}$ in Eq.~\ref{eq:general_t} 
but also depend on neutrino energies $E_{\nu}$ due to 
kinematics of $\pi$-decay and the redshift energy loss. 
$z_{up}$ is given by,
{\footnotesize
\begin{equation}
1+z_{up}\! =\! \left\{
\begin{array}{l}
1\;\;\;\;\;\;\;\;\;\;\;\;\;\;\;\;\;\;\;\;\;\;\;\;\;\;\;\;\;\;\;\;\;\;\;\;\;\;\;\;\;\;
x_1E_{\Delta}x^{+}_R(1-r_\pi)\leq E_\nu,\\
\left(\!{x_1E_{\Delta}\over E_\nu}x^{+}_R(1\!\!-\!\!r_\pi)\!\right)^{1\over 2}\\ 
\;\;\;\;\;\;\;\;\;\;\;
{x_1E_{\Delta}x^{+}_R(1-r_\pi)\over (1+z_{max})^2} \!\leq\! 
  E_\nu \!\leq\! x_1E_{\Delta}x^{+}_R(1-r_\pi) ,\\
1+z_{max}\;\;\;\;\;\;\;\;\;\;\;\;\;\;\;\;\;\;\;\;\;\;\;\;\;\;\;\;\;\;\;
 E_\nu \leq {x_1E_{\Delta}x^{+}_R(1 - r_\pi) \over (1+z_{max})^2}. \\
\end{array} \right.
\label{eq:redshift_up}
\end{equation}
}
$z_{down}$ is also given by Eq.~\ref{eq:redshift_up} replacing $x^+_R$ by $x^-_R$.

See appendix \ref{append:constant} for the case of the astronomical objects 
of which cosmological evolution become constant above a certain redshift (See Ref.~\cite{hopkins2006} for example).

\section{\label{subsec:validity}Validity of the analytical formula}
The analytical formula for estimating cosmogenic neutrino 
fluxes (Eqs.~\ref{eq:gzk_final}, \ref{eq:gzk_redshift_final} and \ref{eq:redshift_up}) 
is derived under several assumptions.
Here we demonstrate the applicability of the formula in 
estimating neutrino flux in 100~PeV $\lesssim E_\nu \lesssim$ 10~EeV which is 
the main energy range of several cosmogenic neutrino searches~\cite{icecubeEHE2011,icecubeEHE2010}. 
In Table~\ref{table:fluxes}, the cosmogenic neutrino integral flux
above 1~EeV obtained by the analytical formula with $\alpha=2.5$ are presented.
We use the UHECR intensity $F_{\rm CR}(\geq E_{\rm GZK}) = 2.96\times 10^{-21}\ {\rm cm}^{-2}\ \sec^{-1} {\rm sr}^{-1}$ 
in the present study, which is obtained from the measurement of the HiRes experiment~\cite{hires04}.
The fluxes obtained by the full numerical calculations with the same or comparable source 
evolution parameters are also listed for comparison.
The values in each parameter subset show an agreement within a factor of two for a comparable evolution scenario
in the wide range of parameter numbers.

\begin{figure}[tb]
  \includegraphics[width=0.4\textwidth]{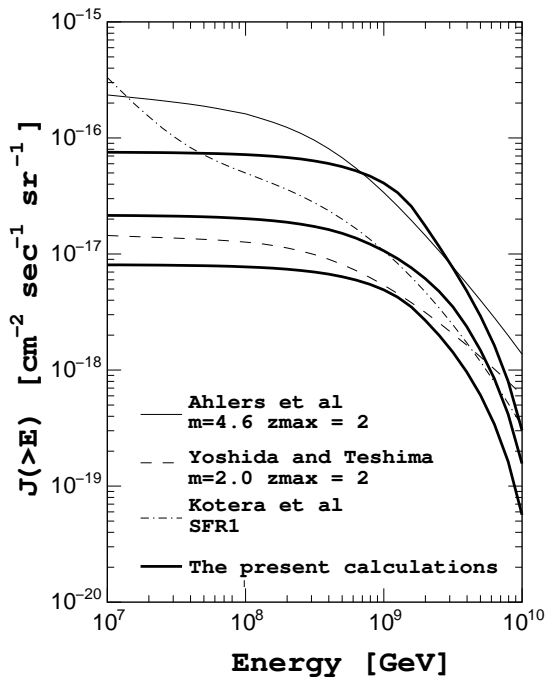}
  \caption{Integral neutrino fluxes, $J$ [cm$^{-2}\ \sec^{-1}$ sr$^{-1}$], as a function of neutrino energy. 
    Bold lines represent the present analytical estimates and thin lines represent corresponding predictions by the full numerical calculations~\cite{ahlers2010} or the Monte-Carlo simulations~\cite{yoshida93,kotera2010}
\label{fig:gzk_fluxes_comparison}} 
\end{figure}

Figure~\ref{fig:gzk_fluxes_comparison} presents the neutrino fluxes obtained with 
the present analytical estimation and the full blown numerical calculations.
The fluxes calculated with the different techniques show the best agreement at $\sim$1~EeV, the central energy 
in the cosmogenic neutrino search with IceCube~\cite{icecubeEHE2011}.
The present formula provides a reasonable 
estimate of the neutrino flux from 100~PeV to 10~EeV
with uncertainty of factor of $\sim$two. 
Some deviations in the analytical formula from the full blown numerical calculations arise mainly from
the uncertainty in the intensity of the extra-galactic UHECR component allowed by the observed
UHECR spectrum, and the accuracy of the approximations used in derivation of the analytical formula.
We discuss these issues in Sec.\ref{sec:discussion}.

\begin{table}[t]
\begin{center}
\begin{tabular}{lc}
\hline
\hline
$\nu$ Flux Model & Integral Flux\\
                 & F(${\rm E}_{\nu}\geq 1$ EeV) [cm$^{-2}\ \sec^{-1}$ sr$^{-1}$] \\
\hline
\hline
Yoshida and Teshima~\cite{yoshida93} &    \\
$m = 2.0,z_{max}=2.0$& $5.39\times 10^{-18}$ \\
Ahlers {\it et al.}~\cite{ahlers2010} &    \\
$m = 2.0,z_{max}=2.0$& $1.85\times 10^{-18}$ \\
("the minimal case") & \\
The analytical formula &    \\
$m = 2.0,z_{max}=2.0$& $4.91\times 10^{-18}$ \\
\hline
Kotera {\it et al.}~\cite{kotera2010} &     \\
SFR1 & $1.07\times 10^{-17}$ \\
The analytical formula &    \\
$m = 3.4 (z\leq 1.0)$ &  \\ 
const. $(1\leq z \leq 4)$ & $1.07\times 10^{-17}$ \\
\hline
Ahlers {\it et al.}~\cite{ahlers2010} &    \\
$m = 4.6,z_{max}=2.0$& $3.39\times 10^{-17}$ \\
("the best fit") & \\
The analytical formula &    \\
$m = 4.6,z_{max}=2.0$& $4.09\times 10^{-17}$ \\
\hline
Kalashev {\it et al.}~\cite{kalashev02} &    \\
$m = 5.0,z_{max}=3.0$& $7.38\times 10^{-17}$ \\
The analytical formula &    \\
$m = 5.0,z_{max}=3.0$& $8.42\times 10^{-17}$ \\
\hline
Kotera {\it et al.}~\cite{kotera2010} &    \\
FR II & $6.74\times 10^{-17}$ \\
The analytical formula &    \\
$m = 5.02 (z\leq 1.5)$ & \\
const. $(1.5\leq z \leq 2.5)$ & $5.21\times 10^{-17}$ \\
\hline
\hline
\end{tabular}
\caption{Cosmogenic neutrino fluxes predicted by the model-dependent full numerical calculations and those given
by the present analytical formula with the corresponding parameters on source evolution. 
The numbers by the full calculations were converted to be the sum over all three neutrino flavors
from the original when appropriate.
}
\label{table:fluxes}
\end{center}
\end{table}

\begin{figure}[h]
  \includegraphics[width=0.4\textwidth]{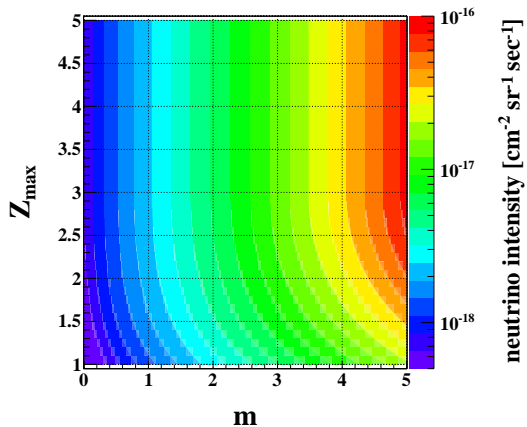}
   \caption{Integral neutrino fluxes with energy above 1~EeV, $J$~[cm$^{-2}\ \sec^{-1}$ sr$^{-1}$], 
    on the plane of the source evolution parameters, $m$ and $z_{max}$.
   \label{fig:gzk_fluxes}}
\end{figure}

\section{\label{sec:result}Results}
\subsection{\label{subsec:relation}
The relation between the $\nu$ flux and the cosmological evolution of the sources}
Shown in Fig.~\ref{fig:gzk_fluxes} is the distribution of the cosmogenic neutrino integral 
fluxes with energies above 1~EeV in the parameter space of the evolution of UHECR sources 
($m,z_{max}$) calculated using the derived analytical formula.
The fluxes vary by more than
an order of magnitude with the evolution parameters. 
The distribution demonstrates that the neutrino intensity can indeed be an observable 
to imply the characteristics of the UHECR sources.
The plot shows that cosmogenic neutrino flux around 1~EeV is mostly determined by source 
emissivity history up to redshift of $z_{s}\sim 3$. 
This is because the contributions of sources
at $z_{s}\gtrsim 3$ represent only a small fraction of the total flux 
due to the redshift dilution~\cite{kotera2010}.

\subsection{\label{sec:constraints} Constraints on UHECR origin with the IceCube diffuse neutrino flux limit}

\begin{figure*}[bt]
  \includegraphics[width=2.0in]{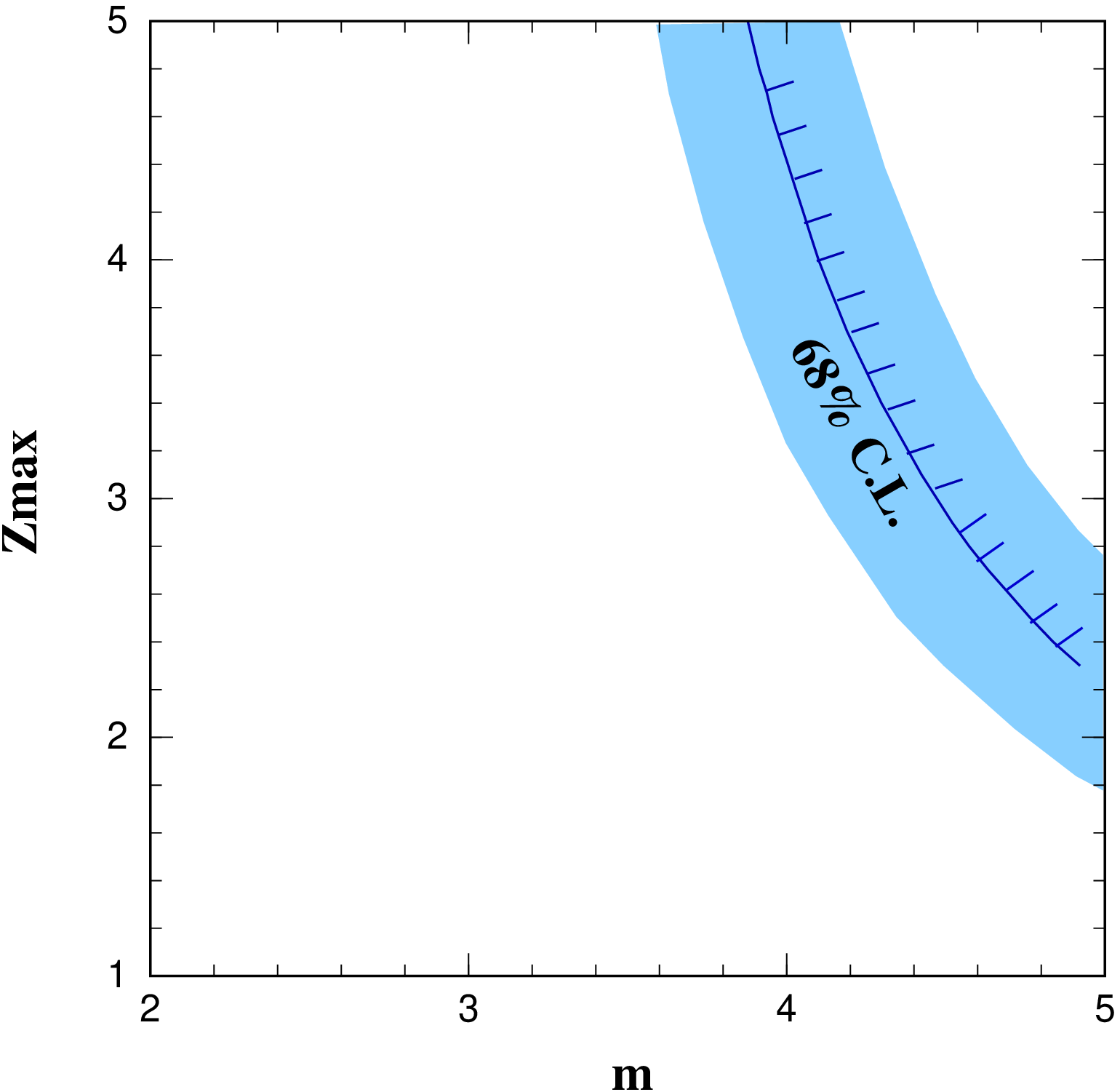}
  \includegraphics[width=2.0in]{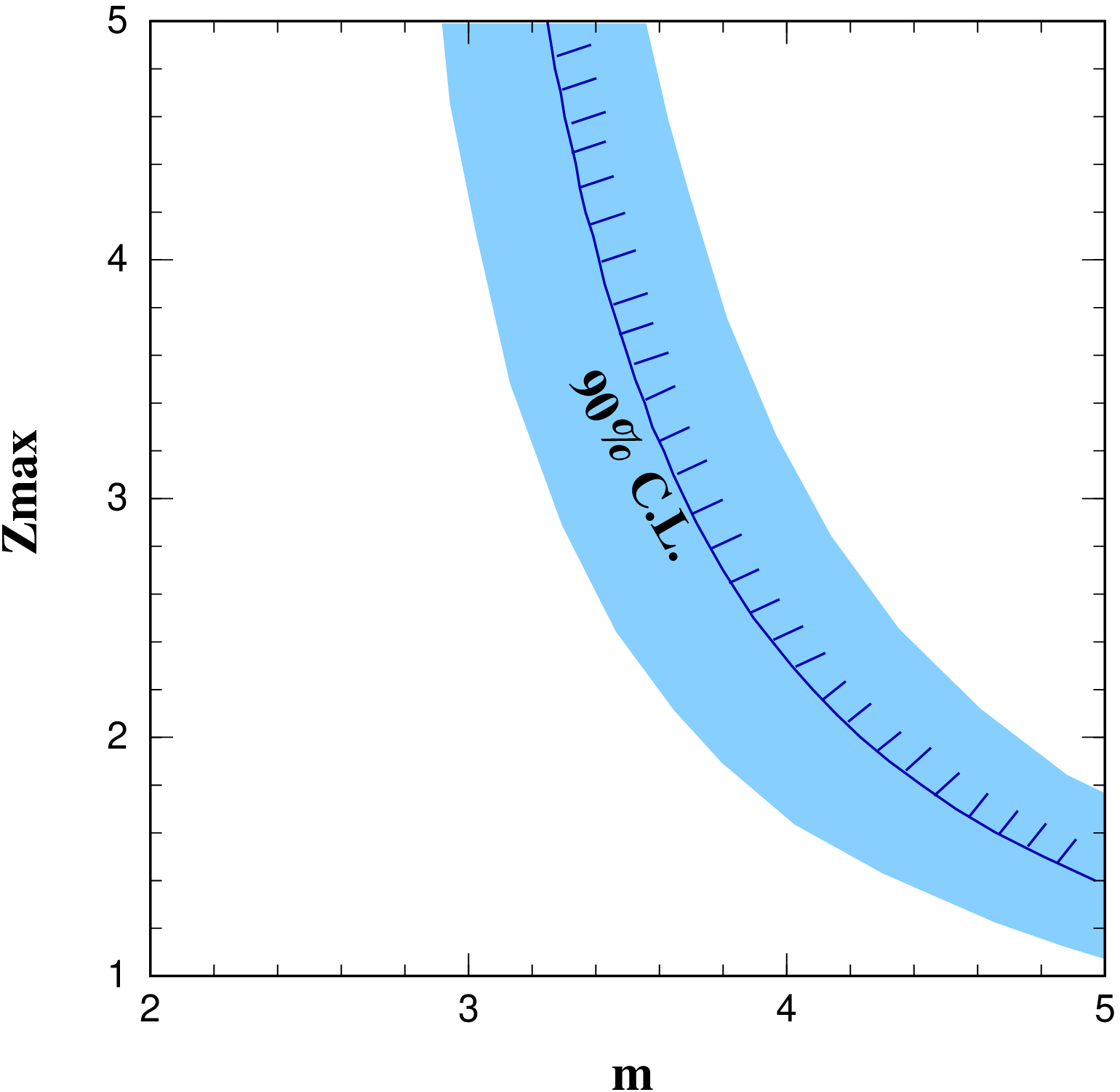}
  \includegraphics[width=2.0in]{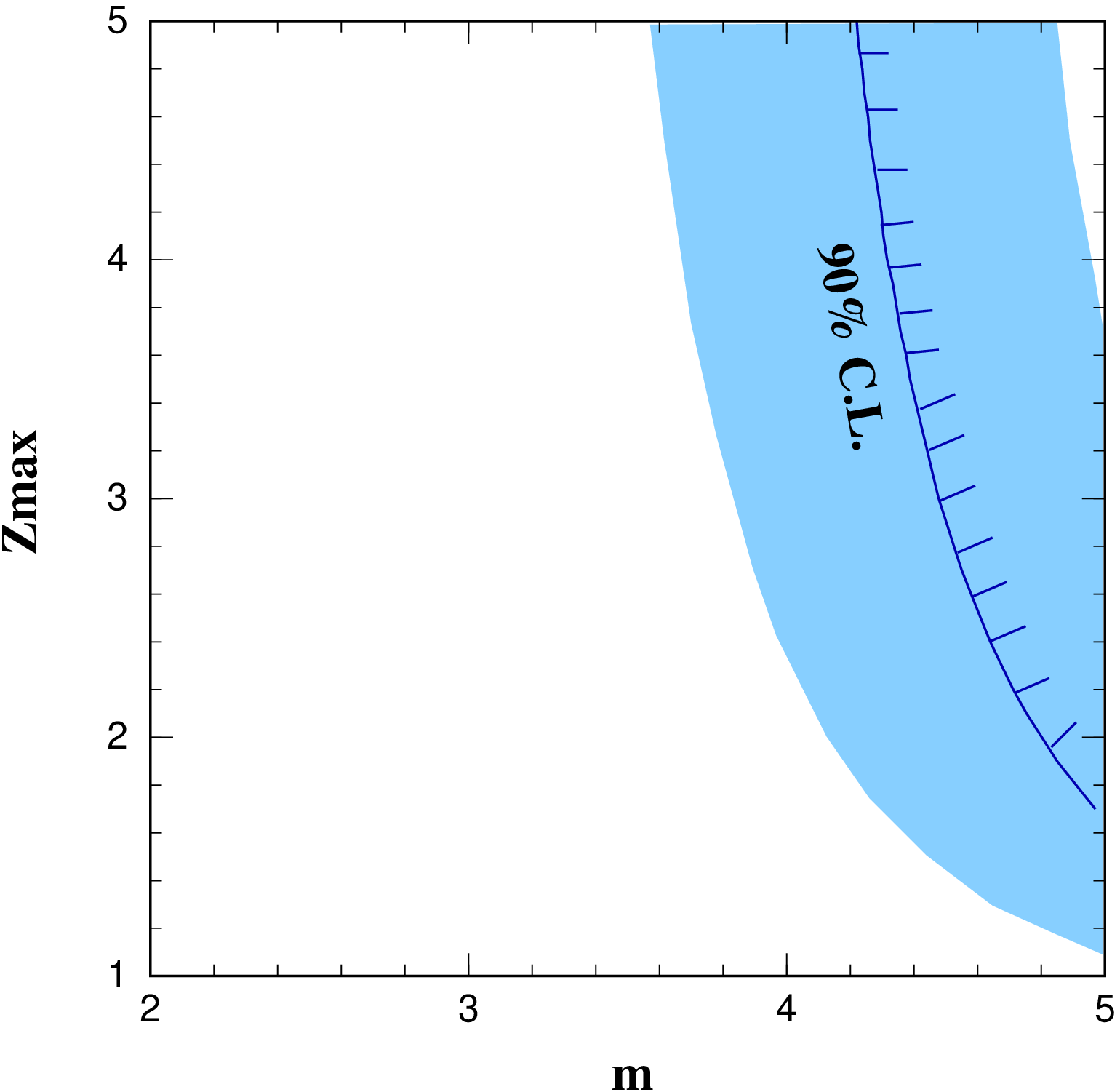}
  \caption{Constraints on the UHECR source evolution $m$ and $z_{max}$ 
    with the IceCube 2008--2009 flux limit~\cite{icecubeEHE2011} (left)
    and with the full IceCube five-year sensitivity~\cite{aya2011} (middle).
    The area above the solid lines are excluded by null detection of $\nu$ events.
    The shaded belts represent uncertainties in the present analytical 
    estimation. The right panel shows the full IceCube 5 year constraint
    when the emission rate per co-moving volume becomes constant above $z_{s}$ of 1.0.
    Excluded region at 68 \% confidence level (corresponding to $\simeq 1.1$ events
    assuming the same background rate of the IceCube 2008--2009 measurement~\cite{icecubeEHE2011}), 
    and 90 \% confidence level ($\simeq 2.2$ events) are displayed. 
 \label{fig:gzk_constraint}}
\end{figure*}

Here we estimate the expected event rates with the IceCube neutrino observatory
by using the derived analytical formula. The analytical function is valid in
the IceCube cosmogenic neutrino detection energy range distributed
around 1~EeV~\cite{icecubeEHE2011}. Convolution of Eq.~\ref{eq:gzk_final} 
with the IceCube neutrino effective area~\cite{icecubeEHE2011,aya2011}
gives the event rate for the entire phase space of the evolution parameter
$m$ and the maximal redshift $z_{max}$. Full mixing in the standard 
neutrino oscillation scenario is assumed and the intensity of neutrinos of
each of three neutrino flavors corresponds to one third of
the estimated neutrino intensity by the analytical function.
The Feldman-Cousins upper bound~\cite{feldman98}
then defines the excluded region on the $m$-$z_{max}$ plane at a given 
confidence level. Figure~\ref{fig:gzk_constraint} displays the resultant
constraints. The shaded region represents the factor of two uncertainty
in the analytical estimation discussed in the previous section.
The upper-limit with the IceCube 2008--2009 data~\cite{icecubeEHE2011} has already started to
constraint on hypotheses of UHECR sources with strong evolution of $m\gtrsim 4.5$. While this bound
may be still weaker than that by the Fermi diffuse $\gamma$-ray flux
measurement~\cite{ahlers2010}, nevertheless the limit by neutrinos is important because the neutrino estimate
does not involve the uncertainties of the assumptions of $E_{\rm max}$
nor the EBL intensity.
The full IceCube five-year observation would certainly probe the most interesting
region of the source evolution phase space where the strong candidates for the UHECR sources of the powerful
astronomical objects such as radio galaxies and GRBs are included.

\section{\label{sec:discussion} Discussion}
The derived analytical formula to calculate intensities of the neutrinos produced by the GZK mechanism
in the range between 100~PeV and 10~EeV is used to constrain the cosmological evolution of the UHECR sources.

The largest uncertainty in the present analytical formula at the lower energy
range ($E_\nu$ $\ll$ 1~EeV) is due to the omission of the IR contribution
to the cosmogenic neutrino production. 
Photo-produced pions from the UHECR interactions
with the IR background are major origin of neutrinos with energies below 10~PeV,
however, they are relatively minor in the higher energies where we mainly discuss
in the aspect of the cosmogenic neutrino detection by IceCube. 
The amount of the IR contribution was studied, for example, in the calculations in Refs.~\cite{kotera2010, ahlers2010}. 
The study in the former reference exhibits much higher contribution of the IR background than in the latter where the effect
is suppressed mainly due to the introduction of the minimal energy of extra-galactic UHECR population. 
These differences can be seen
in figure~\ref{fig:gzk_fluxes_comparison}; the low energy component in Ref.~\cite{kotera2010}
is substantially emphasized compared to the other calculations.
These variations in the estimation of the IR contribution to the cosmogenic neutrino intensities 
are considered to be an additional uncertainty to the IR background yield itself 
which is also not firmly understood~\cite{takami09,COBE-Firas}.
Here we would like to emphasize that the omission of 
the IR background leads to a conservative constraint on the UHECR
source evolution. 

The second largest uncertainty is concerned with $F_{\rm CR}$, the UHECR intensity above
$E_{\rm GZK}\simeq 10^{20}$~eV. The works in Refs.~\cite{ahlers2010,ahlers2011} allowed a sizable variation
in the UHECR intensity within 99\% confidence level of the statistical test against the observed data. 
It indicates that an extreme case of the UHECR intensity may lead to a large departure 
from the present estimate of the neutrino fluxes.
For instance, the difference of their estimate for the scenario of $(m, z_{max})=(2,2)$ found in 
Table~\ref{table:fluxes} arises from their assumption of very steep UHECR spectrum
leading to a minimal $F_{\rm CR}$ estimation. This uncertainty is, however, expected to
be reduced in future when the statistical uncertainties in the observations of UHECRs and$/$or
the systematic uncertainty in the energy estimation are improved.

We would like to also emphasize that the neutrino intensity
below 10~EeV is not largely affected by the detailed behavior of UHECR proton propagation
in extra-galactic space. This is because these neutrinos are mostly generated at cosmological
distances away, which are substantially longer than the UHECR proton 
energy attenuation length in the CMB field.
It is also suggested by no explicit dependence of $\lambda_{\rm GZK}$ in the final formula Eq.~\ref{eq:gzk_final}.
This is related with the fact that the cosmogenic
flux below 10~EeV is relatively insensitive to $E_{\rm max}$ and $\alpha$, the maximal injection energy
of UHECR protons from their sources and the spectral index of UHECR spectrum, respectively, 
while the flux above 10~EeV is sensitive to those parameters~\cite{yoshida93,kalashev02,takami09}.
A scan of the parameter spaces of the cosmogenic
neutrino sources for some known classes of astronomical objects
with a numerical Monte-Carlo method was made in Ref.~\cite{kotera2010,takami09} 
and it was also shown that
the intensity around 1~EeV is stable against $E_{\rm max}$ variation and the transition models 
between the Galactic and extra-galactic cosmic-ray components.
%
%
These observations are consistent with
the fact that the neutrino intensities around 1 EeV by the relatively
old works~\cite{yoshida93,ESS} assuming harder UHECR spectrum of $\alpha=2.0$
and higher $E_{\rm max}$, and those by the recent works~\cite{ahlers2010,kotera2010} with $\alpha\sim2.5$ 
shows an agreement also within a factor of two. 
The difference between the present analytical formula and the full blown simulation
above $\sim$10~EeV in figure~\ref{fig:gzk_fluxes_comparison} is
attributed to responses to $E_{\rm max}$.
The present analytical estimates
of neutrino fluxes for 100~PeV $\lesssim E_\nu \lesssim$ 10~EeV, the main energy range
by the IceCube cosmogenic neutrino search, is robust against these parameters.
We should note however that we use $F_{\rm CR}$
for the normalization constant assuming $E_{\rm max}\gg E_{\rm GZK}$.
If $E_{\rm max}$ is comparable or lower than $E_{\rm GZK}$, the neutrino yield
strongly depends on $E_{\rm max}$ and the present simple treatment is not capable
of providing reasonable estimates of the cosmogenic neutrino fluxes.

The present analysis indicates that a five-year observation by the IceCube
observatory will scan the source evolution parameter space of the most interest
where many of the proposed UHECR astronomical sources are distributed.
A null neutrino observation then would imply that either UHECR sources are only 
locally distributed ($z_{max} \lesssim 1$), very weakly evolved ($m\lesssim 3$), or
the UHECRs are not proton dominated but heaver nuclei such as irons after all.
The first two possibilities may lead to a speculation about
the highest energy particle emission from
an entirely different and probably dimmer class of objects than currently suggested.
The last possibility has also been discussed with the measurement of
the depth of maximum of air-showers by the Auger collaboration~\cite{augerXmax2010}.
Neutrino search in ultra-high energies provides a complementary constraint
on the proton fraction of UHECRs in this case~\cite{ahlers2009}.

\section{\label{sec:summary} Summary}
We have derived the analytical formula to estimate the cosmogenic
neutrino fluxes for wide range of cosmological evolution parameters of UHECR emission sources.
The analytical formula provides a practical tool for estimating the neutrino intensity at
around EeV energy region with a limited accuracy within a factor of $\sim$two. 
The obtained analytical estimates have indicated
that the present IceCube neutrino limit in 100~PeV -- 10~EeV energies
disfavors the scenarios with the strongly-evolved UHECR sources. The future IceCube observation
will be able to scan most of the interesting parameter space of UHECR source evolution.
Furthermore, while the deep and highly energetic part of Universe is inaccessible with 
photons or cosmic-rays due to the CMB field, the current study implies that the neutrinos 
can be used as a rare tool to probe the far Universe.


With the greater statistics of ultra-high energy neutrino detections by the future neutrino telescopes
of $\sim$100~km$^2$ areas, such as ARA~\cite{ARA} and ARIANNA~\cite{ARIANNA}, 
the analytical formula allows us to specify the astronomical classes of ultra-high energy cosmic-ray sources.
The pioneer prediction of the cosmogenic neutrinos
in 1960's~\cite{berezinsky69} will finally lead to revealing the characteristics of UHECR emission mechanism
in a near future.

\begin{acknowledgments}
We wish to acknowledge Kumiko Kotera
for useful discussions in preparations for this work.
The authors are also grateful to Markus Ahlers and Spencer Klein
for valuable comments on the manuscript. 
A.~Ishihara acknowledges support by the Research Fellowships of the Japan Society 
for the Promotion of Science for Young Scientist.
This work was supported in part by
the Grants-in-Aid in 
Scientific Research from the JSPS in Japan.
\end{acknowledgments}


\appendix
\section{\label{append:energetics} The flux calculation based on energetics}

Recently the possible upper-limit on the cosmogenic neutrino flux has been discussed
solely using the Fermi measurement on extra-galactic diffuse $\gamma$-ray background~\cite{wang2011}.
In this work, the neutrino flux was approximately estimated by the energetics argument;
calculating energy channeling into secondary neutrinos from a UHECR proton during its propagation in the CMB field.
The neutrino flux is then calculated by
\begin{eqnarray}
E_{\nu}{dJ_{\nu}\over dE_{\nu}}&=& n_0 c\int\limits^{z_{max}}_0 dz_{s}
\psi(z_{s})\left|{dt\over dz}(z_s)\right| \nonumber\\
 &&\int dE_{\rm CR} E_{\rm CR}{dN_{\rm CR}\over dE_{\rm CR}} R_\nu(E_{\rm CR}) {d\rho_\nu\over dE_\nu}. 
\label{eq:general_energetics}
\end{eqnarray}
Here $dN_{\rm CR}/dE_{\rm CR}$ is the injected UHECR proton spectrum ($\sim E_{p}^{-\alpha}$) at the source redshift $z_{s}$,
$R_\nu$ is a fraction of UHECR proton injection energy carried by the secondary neutrinos, 
$d\rho_\nu/dE_\nu$ is a distribution of neutrino energy. $R_\nu$ was calculated in the earlier work~\cite{ESS}
represented by a numerically fitted function as 
\begin{equation}
R_\nu = {0.45\over 1+ \left({2\times 10^{11}\ {\rm GeV}\over (1+z_{s})E_{\rm CR}}\right)^2}
\label{eq:energy_fraction}
\end{equation}
While the original work~\cite{wang2011} represented $d\rho_\nu/dE_\nu$ as $\sim \delta(E_\nu-E_{\rm CR}/(20(1+z_{s})))$
approximating each secondary neutrino receiving 1/20 of the UHECR proton energy,
we found that the single pion kinematics approximation would give a better agreement 
in the neutrino spectral shape with those obtained by the full blown simulation. It is then written as
\begin{equation}
{d\rho_\nu\over dE_\nu}\simeq (1+z_{s})[E_{\rm CR}(x_R^+-x_R^-)(1-r_\pi)]^{-1}\ln\left({x_R^+\over\xi_R}\right),
\label{eq:energy_distribution}
\end{equation}
where $\xi_R$ is given by Eq.\ref{eq:xi_range}, replacing $E^{g}_{\nu}$ with $E_\nu(1+z_{s})$.

This approach has an advantage that it does not rely on the $\Delta$-resonance approximation.
Although we are not able to find out a complete analytical solution 
of the integrals in Eq.~\ref{eq:general_energetics},
the numerical calculations indeed confirmed that the formula Eq.~\ref{eq:general_energetics}
reasonably reproduces the full simulation/numerical calculation results. It gives a better
agreement than our formula at around 100~PeV, owing to inclusion of direct pion production
yielding a pair of $\pi^+\pi^-$ by Eq.~\ref{eq:energy_fraction}~\cite{ESS}.
However, this energetics-based formulation significantly overestimates 
neutrino intensities with energy above 1~EeV.  We suspect it due to the neglecting of the energy loss
of UHECR protons. Energy of an UHECR proton is in many cases largely lower 
than its injected energy when it yields neutrinos,  because of energy loss by the photo-pion production 
during the UHECR propagation. Without accounting this effect,
higher energy neutrino production is over weighted in the formulation.
The overproduced high energy neutrinos are then red-shifted and accumulated even in PeV regime
when UHECR sources are strongly evolved. As a consequence, the estimated intensity departs 
from the calculation with the full blown simulation in case of the strong evolution scenario.
Since the GZK neutrino search by the IceCube is sensitive to EeV range and emission 
from strongly evolved sources, we concluded that the energetics-based formulation
is not appropriate for our purpose.

\section{\label{append:constant} The analytical formula for the partially constant source evolution}
Some classes of astronomical objects, like the galaxy star formation, 
seem to exhibit evolution nearly constant above a certain redshift.
For these cases, cosmological evolution of sources are written as, 
$\psi(z_{s})\sim (1+z_{s})^m$ for $0\leq z_{s}\leq \tilde{z}$
and $\psi(z_{s})\sim (1+\tilde{z})^m$ for $\tilde{z}\leq z_{s}\leq z_{max}$.
The redshift dependence term $\zeta$ (Eq.~\ref{eq:gzk_redshift}) is then obtained with minor modifications
on the formula Eq.~\ref{eq:gzk_redshift_final} and given as,
\begin{equation}
\zeta = \zeta_{\tilde{z}} + \tilde{\zeta}
\label{eq:gzk_redshift_const}
\end{equation}
where $\zeta_{\tilde{z}}$ is given by Eq.~\ref{eq:gzk_redshift_final} with
replacing $z_{max}$ by $\tilde{z}$, accounting for the evolution up to $z_{s}\leq \tilde{z}$.

The additional term $\tilde{\zeta}$ is obtained in the similar functions as Eq.~\ref{eq:gzk_redshift_final} as,
{\small 
\begin{eqnarray}
\tilde{\zeta} &=& e^{-2}{1\over \gamma_\alpha}\Omega_M^{-{\alpha\over 3}}(1+\tilde{z})^m \nonumber\\
&&\left\{(\Omega_M(1+\tilde{z}_{up})^3+\Omega_{\Lambda})^{\gamma_\alpha\over 3}
\left(x_0^{-(\alpha+1)}\ln{\left({x^{+}_R\over x^{-}_R}\right)}+ \right.\right. \nonumber\\
&& \left.\left. x_1^{-(\alpha+3)}e^{-{1\over x_1}}
\left[\ln\left({x_1{E_{\Delta}\over E_\nu(1+\tilde{z}_{up})^2}x^{-}_R(1-r_\pi)}\right)\right.\right.\right. \nonumber\\
&& \left.\left.\left. +{2\over \gamma_\alpha}\right]\right) -
(\Omega_M(1+\tilde{z}_{down})^3+\Omega_{\Lambda})^{\gamma_\alpha\over 3}x_0^{-(\alpha+1)}
\ln{\left({x^{+}_R\over x^{-}_R}\right)}\right.\nonumber\\
&& \left. -(\Omega_M(1+\tilde{z}_{down})^3+\Omega_{\Lambda})^{\gamma_\alpha\over 3}
x_1^{-(\alpha+3)}e^{-{1\over x_1}}\right.\nonumber\\
&& \left. \left[\ln\left({x_1{E_{\Delta}\over E_\nu(1+\tilde{z}_{down})^2}x^{-}_R(1-r_\pi)}\right) +
{2\over \gamma_\alpha}\right]\right\}.\\\nonumber
\label{eq:gzk_redshift_const_final}
\end{eqnarray}
}

Where $\gamma_\alpha \equiv \alpha-{3\over 2}$ and the redshift bounds $\tilde{z}_{up}$ is given by 
{\footnotesize
\begin{equation}
1+\tilde{z}_{up}\! =\! \left\{
\begin{array}{l}
1+\tilde{z}\;\;\;\;\;\;\;\;\;\;\;\;\;\;\;\;\;\;\;\;\;\;\;\;\;\;\;\;\;\;\;\;\;\;
{x_1E_{\Delta}x^{+}_R(1-r_\pi)\over (1+\tilde{z})^2}\leq E_\nu,\\
\\
\left(\!{x_1E_{\Delta}\over E_\nu}x^{+}_R(1\!\!-\!\!r_\pi)\!\right)^{1\over 2}\\ 
\;\;\;\;\;\;\;\;\;\;
{x_1E_{\Delta}x^{+}_R(1-r_\pi)\over (1+z_{max})^2} \!\leq\! 
  E_\nu \!\leq\! {x_1E_{\Delta}x^{+}_R(1-r_\pi)\over (1+\tilde{z})^2},\\
\\
1+z_{max}\;\;\;\;\;\;\;\;\;\;\;\;\;\;\;\;\;\;\;\;\;\;\;\;\;\;\;\;\;
 E_\nu \leq {x_1E_{\Delta}x^{+}_R(1 - r_\pi) \over (1+z_{max})^2}. \\
\\
\end{array} \right.
\label{eq:redshift_up2}
\end{equation}
\vspace{-0.1cm}

}
$\tilde{z}_{down}$ is written as the same Eq.~\ref{eq:redshift_up2} with replacing 
$x^+_R$ by $x^-_R$.


\end{document}